# Description of the RHIC Sequencer System

T. D'Ottavio, B. Frak, J. Morris, T. Satogata and J. van Zeijts *
Brookhaven National Laboratory, Upton, NY 11973, USA

Abstract

The movement of the Relativistic Heavy Ion Collider (RHIC) through its various states (eg. injection, acceleration, storage, collisions) is controlled by an application called the Sequencer. This program orchestrates most magnet and instrumentation systems and is responsible for the coordinated acquisition and saving of data from various systems. The Sequencer system, its software infrastructure, support programs, and the language used to drive it are discussed in this paper. Initial operational experience is also described.

## 1 INTRODUCTION

Automating collider operations using a sequencer is not a new idea. Fermilab has built and used a sequencer for years [1], and CERN has a sequencer in use for LEP.

When RHIC was built at BNL, the construction of a sequencer was never in doubt. A prototype sequencing program was first used when RHIC was turned on and commissioned in the year 2000. Since then, a major effort has gone into building and using a sequencing system for both RHIC and injector operations at BNL.

Expectations for the sequencing system were what one would expect from any automation tool – improvements in execution times, minimization of errors, consistent playback of procedures, and easier and more automated diagnostics.

This paper will describe the overall sequencing infrastructure that is now in place and how the various pieces are used to sequence procedures at BNL. Additional information on the design and development of the main sequencer GUI pictured in Figure 2 has recently been published [2].

## 2 SYSTEM DESCRIPTION

An overview of the different pieces of the sequencing system is shown in Figure 1. The system is comprised of the sequences, sequence creation and editing tools, two GUIs that allow users to see the possible sequences and run them, support servers which run specialty code and allow other applications to run sequences, and a message logging system used to record sequence progress and aid in diagnostics.

**Sequences, Language and Editing.** The sequences are the heart of the system – the common link that ties all of the pieces together. Sequences describe the tasks to be performed in a linear sequence of steps. Each line in the sequence is either a primitive task (i.e. set a value, trigger an event, etc.) or a call to run a different sequence. This nesting capability adds a great deal of flexibility to the system and has led to a more modular and smaller set of needed sequences. See the "ramping up" example in Figure 2.

Sequences are ASCII files and, therefore, can be edited by any text editor if desired. An alternative method, using a program called TreeBuilder, permits point and click construction of a sequence by selecting from a list of available tasks and input variables for each task.

**Sequencer GUIs.** Parallel development of a GUI to run sequences by the Physics and Controls group led to the development to two sequencer GUIs. The one developed by Physics, called Sequencer (Figure 2) is the main sequencer used to run the RHIC accelerator. It is written in the tcl/tk language and can read and run files in the sequencer language described above. Alternatively, it can issue commands to the Sequencer Server (see below) or run shell scripts. Messages and errors are displayed in an internal message window and stored by the cmlog Server. More information on the Sequencer program is available [2].

A second sequencer GUI called tape (Tool for Automated Procedure Execution) is an alternative C++ based sequencer front-end. It has been used by Controls to run sequences containing custom C++ tasks for items like quench recovery and particle mode switching. Figure 3 shows tape with the quench recovery sequence loaded. Sequences run here can also be run by the main Sequencer via the Sequencer Server if desired.

**Support Servers and Programs.** Two servers were constructed to support the sequencing system – the Sequencer Server and the Launch Server. Both servers use the CDEV protocol [3] to communicate with their clients. The Sequencer Server is dedicated to running commands from the Sequencer GUI, which may use the server to run a custom task written in C++ or a CPU or time intensive task that can reasonably be run in the background.

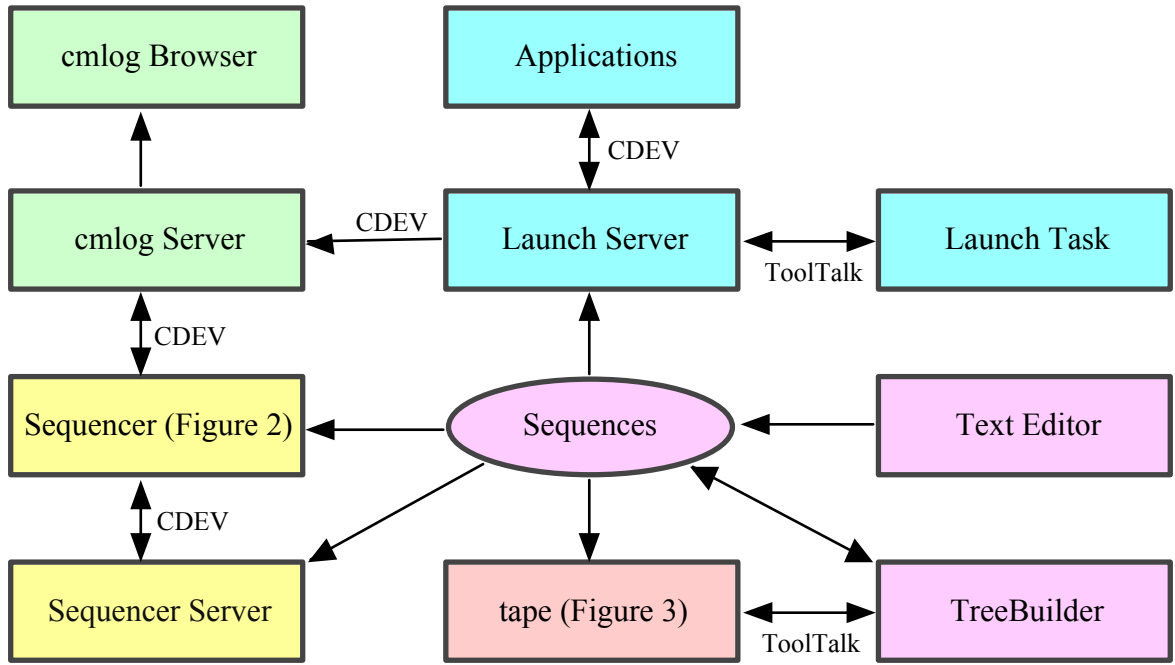

Figure 1: The components of the RHIC Sequencing System include the sequencer GUIs, support servers, sequence creation and editing tools and the sequences. See the text for a description of each component.

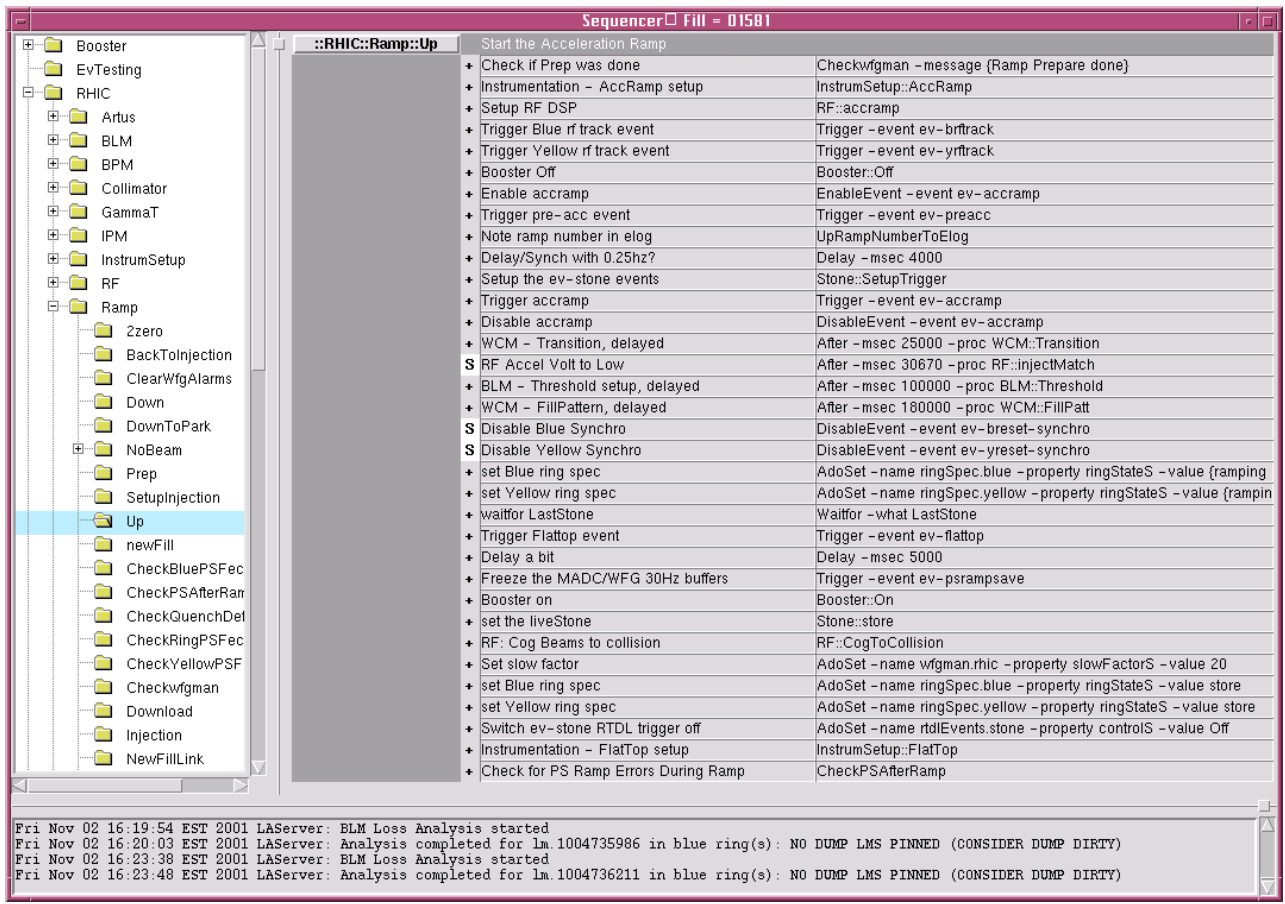

Figure 2: This is the main sequencer used for RHIC operations. It is shown here displaying the RHIC::Ramp::Up sequence, which ramps RHIC from injection energy up to maximum storage energy.

The Launch Server was designed to service requests to run sequences from any client. This gives any application the ability to run a sequence. The server can handle simultaneous requests by spawning off a process to handle a sequence. The ToolTalk communication protocol is used for this purpose.

**Message Logging.** Messages and errors from the Sequencer GUI and from the Launch Server are sent to the cmlog Server [4], where they are stored and available for viewing from the cmlog Browser, which provides custom GUI displays and querying capabilities. The tape program has its own message logging and display system.

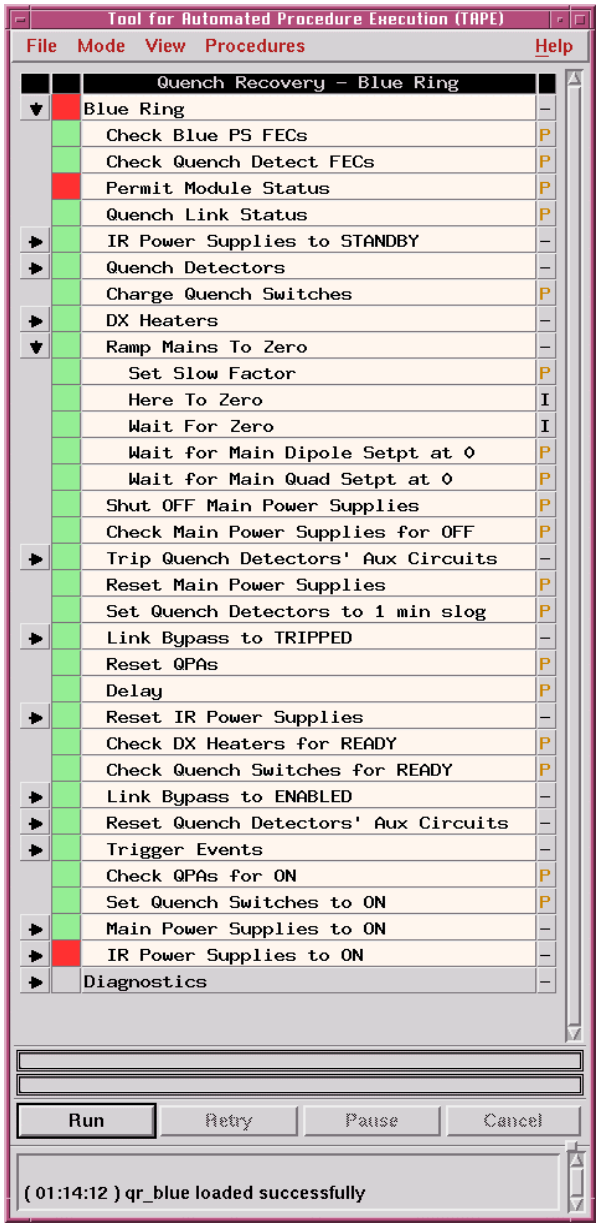

Figure 3: tape is an alternate sequencer GUI shown here after running a quench recovery sequence.

## 3 OPERATIONAL EXPERIENCE

The sequencing system described here has been used heavily during the first year of RHIC operations. As can be seen from the tree of sequences shown in Figure 2, a number of sequences have been created for setting up and checking instrumentation and power supply systems, for ramping and moving through the various states of the collider, and for system recovery and preparation. The sequencing system has been a key contributor to operating RHIC in an efficient and reproducible way and has generally been quite reliable.

One unexpected benefit of developing this system is that it is now being used to automate procedures that have, in the past, been done manually by the Operations staff. In many ways, the sequencer system is now considered a general-purpose automation tool available for a wide variety of automation purposes that might have been handled in the past by custom coding.

## 4 FUTURE WORK

Future work will involve enhancements designed to address limitations in the current system. The sequencer language is currently limited to describing the sequential execution of tasks. Under consideration are enhancements to permit parallel execution and simple if/then/else logic. Code of this nature can currently be executed only within custom code executed on a single line of the sequence.

Enhanced diagnostic capabilities when errors occur during the running of a sequence are also being pursued. Currently, the sequencer GUI highlights a line in the sequence if it encounters an error. The user must then gather the available diagnostic information and determine how to proceed. This process is being automated so that diagnostic information and recommendations are available to the user at any error point in the sequence.